# *Multi-Photon, Multi-Dimensional Hyper-Entanglement using Higher-Order Radix qudits with Applications to Quantum Computing, QKD and Quantum Teleportation*

*Solyman Ashrafi, Logan Campbell*
*NxGen Partners*
*Dallas, TX*
solyman.ashrafi@nxgenpartners.com

**Abstract**—Google recently announced that they had achieved quantum supremacy with 53 qubits (base-2 binaries or radix-2), corresponding to a computational state-space of dimension $2^{53}$ (about $10^{16}$). Google claimed to perform computations that took 200 seconds on their quantum processor that would have taken 10,000 years to accomplish on a classical supercomputer [1]. However, achieving superposition and entanglement of 53 qubits is not an easy task given the environmental noise that decoheres the qubits. In this paper, we claim that one can potentially achieve a similar computational dimension with fewer qudits (not qubits) where each qudit is of a higher radix (greater than 2) using a photonics system (i.e. 16 qudits with radix-10). This paper is a collaborative development between industry and NxGen Partners to explore such an approach. There is a Raytheon technology that uses a Free-Space Optical (FSO) Fabry-Perot Etalon that eliminates the need for adaptive optics [2, 3]. The NxGen technology uses multiple Orbital Angular Momentum (OAM) modes as a new degree of freedom for quantum computing and a multi-dimensional QKD [4-7]. We also claim that the convergence of both broadband, secure communications, quantum computing and quantum teleportation is only possible in a photonics realization. Therefore, the use of photonic qudits allows the extension of security and capacity of the quantum teleportation beyond what was achieved by Chinese Micius quantum satellite [8]. Both the defense and commercial computing industries need such quantum computing systems. A new measure is introduced with computational state-space of dimension, high-fidelity operations, high connectivity, large calibrated gate sets, and circuit rewriting toolchains. This new measure which we call quantum capacity is a practical way to measure and compare progress toward improved system structure of a universal quantum computer.

**Keywords**— *Quantum Computing, Orbital Angular Momentum, higher-radix, qudits, Free-Space Optical (FSO), Quantum Capacity*

## I. INTRODUCTION

Quantum computing has been an area of ongoing research for more than 30 years. It leverages quantum mechanical phenomena to greatly enhance the way in which information is stored and processed. Quantum computing also uses more efficient algorithms than what is possible in classical computing. Although physicists were able to theorize three decades ago how a quantum computer could work, scientists and engineers had difficulty building one. In the last five years, we have seen the hardware and software capability move out of the university labs and into tangible business products. However, the technology still needs to mature for it to become fully enterprise-ready in a way that would deliver meaningful, cost-effective business and military results.

In a quantum computer, the basic unit of information is known as a quantum bit or "qubit." Through quantum mechanical phenomena (superposition and entanglement), these qubits can perform many computations simultaneously in parallel, which theoretically allows the quantum computer to solve a difficult subset of problems much faster than a classical computer.

Quantum technology is still maturing and there are some hurdles left to overcome in order to build fully scalable quantum computers. As just one example, quantum systems are much more sensitive than classical computers to noise. Noise causes a quantum system to decohere and lose its quantum properties. There is a lot of room for progress in devising quantum error correction schemes (also known as fault-tolerant quantum computing), as well as engineering advancements toward suppressing noise effects. Nevertheless, many companies have started looking at quantum computing for both the defense and commercial sectors (i.e. financial, healthcare, manufacturing, media and technology). Governments around the world are forging
ahead with quantum computing initiatives and there are now over 100 companies working on quantum computing and almost all are using qubit as their elementary unit of information.

## II. QUBITS AND HIGHER ORDER RADIX QUDITS

Qubit is the standard unit of information for radix-2, or base-2, quantum computing. The qubit models information as a linear combination of two orthonormal basis states such as the states |0> and |1>. The qubit differs from the classical bit by its ability to be in a state of superposition, or a state of linear combination, of all basis states. Superposition allows quantum algorithms to be very powerful since it allows for parallelism during computation so that multiple combinations of information can be evaluated at once. There are theoretically an infinite number of states for a qubit while in a state of superposition

$$|\Psi\rangle_2 = \alpha|0\rangle + \beta|1\rangle$$

Where $|0\rangle_2 = \begin{pmatrix}1\\0\end{pmatrix}$ and $|1\rangle_2 = \begin{pmatrix}0\\1\end{pmatrix}$

Where α and β are complex numbers and for radix-2

$$\alpha^*\alpha + \beta^*\beta = 1$$

For a higher order radix (i.e. radix-4), we have

$$|\Psi\rangle_4 = \alpha|0\rangle + \beta|1\rangle + \gamma|2\rangle + \delta|3\rangle$$

where

$$|00\rangle_2 = |0\rangle_4 = \begin{pmatrix}1\\0\\0\\0\end{pmatrix}, |01\rangle_2 = |1\rangle_4 = \begin{pmatrix}0\\1\\0\\0\end{pmatrix}$$

$$|10\rangle_2 = |2\rangle_4 = \begin{pmatrix}0\\0\\1\\0\end{pmatrix}, |11\rangle_2 = |3\rangle_4 = \begin{pmatrix}0\\0\\0\\1\end{pmatrix}$$

Where α, β, γ, δ are complex numbers and for radix-4

$$\alpha^*\alpha + \beta^*\beta + \gamma^*\gamma + \delta^*\delta = 1$$

Quantum algorithms need maximal superposition of states. Maximal superposition of 2-state qubit (basis $|0\rangle, |1\rangle$) is

$$\alpha^*\alpha = \beta^*\beta = \frac{1}{2}$$

Maximal superposition of 4-state qudit (basis $|0\rangle, |1\rangle, |2\rangle, |3\rangle$) is

$$\alpha^*\alpha = \beta^*\beta = \gamma^*\gamma = \delta^*\delta = \frac{1}{4}$$

## III. HADAMARD AND CHRESTENSON TRANSFORM

Hadamard gate puts a qubit in maximally superimposed state for base-2 qubit (radix-2):

Hadamard operator $\quad \mathbb{H} = \frac{1}{\sqrt{2}}\begin{bmatrix}1 & 1\\1 & -1\end{bmatrix}$

However, Chrestenson operator does the same thing for a multidimensional qudit.
$\mathbb{C} = multi - dimensional\ Hadamard$ for n-state qudit of radix-n.

Chrestenson $\mathbb{C}$ is an orthogonal matrix with both column and row vectors form an orthogonal set:

$$\mathbb{C}_r = \frac{1}{\sqrt{r}}\begin{bmatrix} w_0^0 & w_0^1 & \cdots & w_0^{r-1} \\ w_1^0 & w_1^1 & \cdots & w_1^{r-1} \\ \vdots & \vdots & \ddots & \vdots \\ w_{r-1}^0 & w_{r-1}^1 & \cdots & w_{r-1}^{r-1} \end{bmatrix}$$

Where

$$w_0 = e^{i\frac{2\pi}{4}\cdot 0} = 1$$
$$w_1 = e^{i\frac{2\pi}{4}\cdot 1} = i$$
$$w_2 = e^{i\frac{2\pi}{4}\cdot 2} = -1$$
$$w_3 = e^{i\frac{2\pi}{4}\cdot 3} = -i$$

or in general

$$w_k = e^{i\frac{2\pi}{r}k}$$

Thus for a radix-4, the Chrestenson is

$$\mathbb{C}_4 = \frac{1}{\sqrt{4}}\begin{bmatrix} 1 & 1 & 1 & 1 \\ 1 & i & -1 & -i \\ 1 & -1 & 1 & -1 \\ 1 & -i & -1 & i \end{bmatrix}$$

and

$$\mathbb{C}_4|0\rangle_4 = \frac{1}{2}\begin{pmatrix} 1 \\ 1 \\ 1 \\ 1 \end{pmatrix} = \frac{1}{2}[|0\rangle + |1\rangle + |2\rangle + |3\rangle] = |b\rangle$$

$$\mathbb{C}_4|3\rangle_4 = \frac{1}{2}\begin{pmatrix} 1 \\ -i \\ -1 \\ i \end{pmatrix} = \frac{1}{2}[|0\rangle - i|1\rangle - |2\rangle + i|3\rangle] = |d\rangle$$

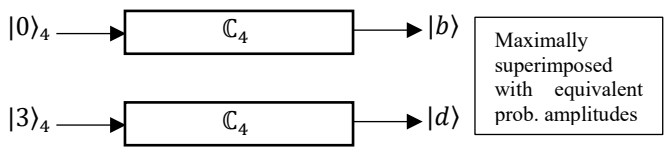

Figure 1. Optical Components for Muxing 4 OAM Modes

## IV. DEFENSE INDUSTRY CONNECTIVITY & COMPUTING

The defense industry requires broadband capabilities possessing the highest level of security in an incredibly complex computing scenario. The bandwidth and security qualities of Free-Space Optics (FSO) make it an attractive technology for military and commercial communications and computing. Photons have a very weak interaction with the environment that makes them perfectly suitable for encoding and transmitting quantum information. However, though photon–photon interactions are challenging when trying to create entangled states, there have been a huge progress in using photonics and multi-photon entanglements. Alternatively, photon-matter interaction can also be used to directly create deterministic gates [6]. There have been even some proposals on the use of graphene that can provide a strong nonlinearity without the technical drawbacks of those atomic systems [9, 6].

FSO communications uses modulated collimated light, usually in the form of an infrared (IR) laser, to transmit data. This affords FSO many appealing qualities such as (i) very high bandwidth capability, (ii) a high level of security through a low probability of detection (LPD) (iii) a low probability of intercept (LPI), and (iv) a signal that is resistant to radio frequency (RF) interference or FCC regulations. FSO also allows us to perform quantum teleportation which is critical for the defense industry. Therefore, we see the convergence of broadband, secure communications, quantum computing and quantum teleportation in a photonics approach.

## V. SAM AND OAM HYPER-ENTANNGLEMENT

The authors from NxGen Partners and Raytheon have been involved in R&D activities leveraging OAM with over 100 patents (issued & pending) and published papers in FSO links [14-37, 2-3], underwater links [38], fiber links [39,40], RF links [41-51] and photon interaction with matter [52,53]. The experience from dealing with OAM states of photons lends itself to the possibility of achieving entangled states using this new degree of freedom.

Spin-Orbital Entanglement

Higher data transmission capacity is one of the primary pursuits in optical communications. Investigation in using different physical properties of a light for data encoding and channel addressing, including amplitude, phase, wavelength, and polarization are some of the approaches. More recently, spatially orthogonal modes and spatial positions have been under intense investigation. A typical method of increasing the transmission capacity in optical communication systems is the multiplexing of multiple independent data channels. For example, multiple independent data channels can be located on different wavelengths, polarizations, or spatial channels, corresponding to wavelength-division multiplexing (WDM), polarization-division multiplexing (PDM), and space-division multiplexing (SDM), respectively.

A special case of SDM is the utilization of orthogonal spatially overlapping and co-propagating spatial modes, known as mode-division-multiplexing (MDM) where, each mode can carry an independent data channel, and the orthogonality enables efficient (de)multiplexing and low inter-modal crosstalk among multiple modes. There are several different types of orthogonal modal basis sets that are potential candidates for such MDM systems. One such set is orbital angular momentum (OAM).

It is well known that a light wave can be interpreted quantum mechanically and thus can be viewed to carry both spin angular momentum (SAM) and OAM. Contrary to SAM (e.g., circularly polarized light), which is identified by the electric field direction, OAM can be interpreted to characterize the "twist" of a helical phase front. Owing to the helical phase structure, an OAM-carrying beam usually has an annular "ring" intensity profile with a phase singularity at the beam center. Depending on the discrete "twisting" rate of the helical phase, OAM beams can be quantified as different states, which are orthogonal while propagating coaxially.

This property allows OAM beams to be potentially useful in improving the performance of optical communication systems. Specifically, OAM states could be used as a different dimension to create an additional set of data carriers in an SDM/MDM system. Importantly, OAM multiplexing does not rely on the wavelength or polarization, indicating that OAM could be used in addition to WDM and PDM techniques to improve system capacity. Compared to other MDM methods, OAM might have some implementation advantages stemming from the circular symmetry of the modes, which make it well-suited for many optical component technologies.

In conjunction with high capacity communications using SAM and OAM, one can use these two properties to entangle them together creating a hyper-entangled photonic state that can be used for multi-dimensional QKD as well as quantum computing.

One way to achieve spin-orbital entanglement is by using metamaterial. The metamaterial could be a dielectric metasurface. These structures are man-made and not found in nature. Metallic metasurfaces produce high loss. Therefore high $n \gg n_0$ dielectric can be used.

SAM-OAM entanglement can also be achieved using q-plates based on liquid crystals. Such an approach can one day produce integrated photonics chips that perform quantum processing. One can use a Geometric Phase Metasurface (GPM) which could be a Si-based component. GPMs are usually used for wavefunction shaping using spin-control. They perform a nano Half-wave plate function that generate a local geometric phase delay. It can have an orientation function $\phi_g = -2\sigma_{\pm}\theta(x,y)$ which defines a geometric phase of light passing through the metasurface at (x, y) for different spin states of a photon $\sigma_{\pm} = \pm 1$ (Right/Left hand circular polarization)

We can design the GPM where

$$\theta(x,y) \rightarrow \theta(r,\phi) = \frac{l\phi}{2}$$

GPM adds or subtracts $\Delta l = \pm 1$, one quanta of OAM depending on sign of spin, but it also flips the spin

$$\sigma_+ = right\ hand\ circular$$
$$\sigma_- = left\ hand\ circular$$

$$|\sigma_+\rangle|l\rangle \xrightarrow{GPM} |\sigma_-\rangle|l + \Delta l\rangle$$

$$|\sigma_-\rangle|l\rangle \xrightarrow{GPM} |\sigma_+\rangle|l - \Delta l\rangle$$

We can start with a Gaussian $l = l_0 = 0$ with linear horizontal (H) polarization

$$|H\rangle|l_0\rangle \rightarrow \frac{1}{\sqrt{2}}[|\sigma_-\rangle|l_0 + \Delta l\rangle + |\sigma_+\rangle|l_0 - \Delta l\rangle]$$

Similarly

$$|V\rangle|l_0\rangle \rightarrow \frac{1}{\sqrt{2}i}[|\sigma_-\rangle|l_0 + \Delta l\rangle - |\sigma_+\rangle|l_0 - \Delta l\rangle]$$

We can then recover the first two Bell states

$$|\Psi^+\rangle = \frac{1}{\sqrt{2}}[|\sigma_+\rangle|l_{-1}\rangle + |\sigma_-\rangle|l_{+1}\rangle]$$
$$|\Psi^-\rangle = \frac{1}{\sqrt{2}}[|\sigma_+\rangle|l_{-1}\rangle - |\sigma_-\rangle|l_{+1}\rangle]$$

Fidelity between recovered density $\tilde{\rho}$ and theoretical $\rho$ is

$$F(\rho,\tilde{\rho}) = Tr\left(\sqrt{\tilde{\rho}^{\frac{1}{2}}\rho\tilde{\rho}^{\frac{1}{2}}}\right)$$

By flipping the GPM
($l\ can\ flip\ sign\ \Delta l = -1) for\ \overline{GPM}\ flipped$

$$|\sigma_+\rangle|l\rangle \xrightarrow{\overline{GPM}} |\sigma_-\rangle|l - \Delta l\rangle$$

$$|\sigma_-\rangle|l\rangle \xrightarrow{\overline{GPM}} |\sigma_+\rangle|l + \Delta l\rangle$$

We can then recover the second two Bell states

$$|\Phi^+\rangle = \frac{1}{\sqrt{2}}[|\sigma_+\rangle|l_{+1}\rangle + |\sigma_-\rangle|l_{-1}\rangle]$$
$$|\Phi^-\rangle = \frac{1}{\sqrt{2}}[|\sigma_+\rangle|l_{+1}\rangle - |\sigma_-\rangle|l_{-1}\rangle]$$

We can also extend this to hyper-entangled states with qudits and in fact construct C-Not gate with Polarization as a control bit and helicity as target bit (we can also do that in reverse) or even beam path as the target bit and combine the methods as shown below:

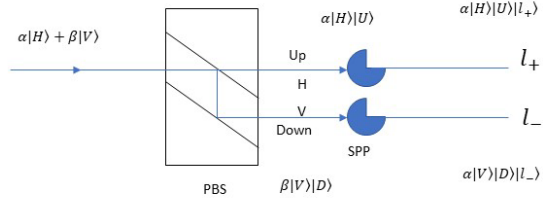

Figure 2. Optical Components for Muxing 4 OAM Modes

$$|0\rangle_n = |H\rangle|U\rangle|l_+\rangle$$
$$|1\rangle_n = |V\rangle|D\rangle|l_-\rangle$$

Where *n* = number of qudits

In general, we can have a hyper-entangled state.

$$|\Psi\rangle_n = \frac{1}{\sqrt{2}}|0\rangle_n - \frac{1}{\sqrt{2}}|1\rangle_n$$

Polarization as control bit, OAM as target

H=1
V=0

$$\alpha|H\rangle|l_+\rangle + \beta|V\rangle|l_-\rangle \rightarrow \alpha|H\rangle|l_+\rangle + \beta|V\rangle|l_+\rangle = (\alpha|H\rangle + \beta|V\rangle)|l_+\rangle$$

$$|H\rangle|l_0\rangle = \frac{1}{\sqrt{2}}[|\sigma_R\rangle + |\sigma_L\rangle]|l_0\rangle$$
$$\rightarrow \frac{1}{\sqrt{2}}[|\sigma_L\rangle|l_{\Delta l}\rangle + |\sigma_R\rangle|l_{-\Delta l}\rangle]$$

Also

$$|V\rangle|l_0\rangle = \frac{1}{\sqrt{2i}}[|\sigma_R\rangle - |\sigma_L\rangle]|l_0\rangle$$

Gives

$$\frac{1}{\sqrt{2}}[|\sigma_L\rangle|l_{\Delta l}\rangle - |\sigma_R\rangle|l_{-\Delta l}\rangle]$$

2-Bell states

$$|\Psi^\pm\rangle = \frac{1}{\sqrt{2}}[|\sigma_R\rangle|l_{-1}\rangle \pm |\sigma_L\rangle|l_{+1}\rangle]$$

Polarization entanglement

$$|\Psi\rangle = \frac{1}{\sqrt{2}}[|H\rangle|V\rangle - |V\rangle|H\rangle]$$

Take $\alpha|H\rangle + \beta|V\rangle$ convert to polarization-path hyper-entangled state

$$\alpha|H\rangle|U\rangle + \beta|V\rangle|D\rangle$$

Then do the OAM entanglement on that

Start with this

$$\alpha|H\rangle|U\rangle + \beta|V\rangle|D\rangle$$

and end with this with helicity entanglement

$$\alpha|H\rangle|U\rangle|l_+\rangle + \beta|V\rangle|D\rangle|l_-\rangle$$

qudits

$$|\Psi\rangle_n = \frac{1}{\sqrt{2}}[|0\rangle_n - |1\rangle_n]$$

Superposition

$$\frac{1}{\sqrt{2}}[|0\rangle \pm e^{i\theta}|1\rangle]$$

Steps with first C-NOT

$$[\alpha|l_+\rangle + \beta|l_-\rangle]|H\rangle \rightarrow \alpha|l_+\rangle|H\rangle + \beta|l_-\rangle V\rangle$$

Second C-NOT

$$\alpha|l_+\rangle H\rangle + \beta|l_-\rangle|V\rangle \rightarrow [\alpha|H\rangle + \beta|V\rangle]|l_+\rangle$$

Then

$$|l_+\rangle \rightarrow |G\rangle$$

And outcome

$$|l_+\rangle|H\rangle \xrightarrow{QW} \frac{|l_+\rangle}{\sqrt{2}}[|H\rangle - i|V\rangle] \xrightarrow{\text{q-plate}} \frac{|G\rangle}{\sqrt{2}}[|H\rangle + i|V\rangle] \xrightarrow{QW} |G\rangle|H\rangle$$

$$|l_-\rangle|V\rangle \xrightarrow{QW} \frac{|l_-\rangle}{\sqrt{2}}[|H\rangle + i|V\rangle] \xrightarrow{\text{q-plate}} \frac{|G\rangle}{\sqrt{2}}[|H\rangle + i|V\rangle] \xrightarrow{QW} |G\rangle|V\rangle$$

For photons, if we only use polarization, we will only present a qubit (radix-2). However, if we leverage OAM, we can present a qudit (higher order radix). Therefore, for a pump beam of zero-angular momentum multi-dimensional entangled state for 2-photons

$$|\Psi\rangle = C_{0,0}|00\rangle + C_{1,-1}|1\rangle|-1\rangle + C_{-1,1}|-1\rangle|1\rangle + C_{2,-2}|2\rangle|-2\rangle + C_{-2,2}|-2\rangle|2\rangle$$

Neither photon in this state has a well-defined OAM after parametric down-conversion. The measurement of 1 photon defines its OAM state and projects the second one into the corresponding OAM state.

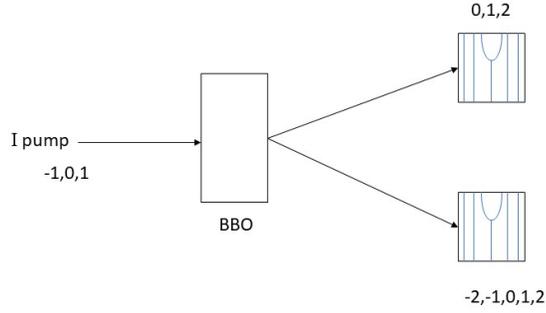

Figure 3. Focusing Optics for Muxing 4 OAM Modes

Two-photon Interferometry

There are two types of Spontaneous parametric Down-conversion (SPDC).

Type I:

Non-linear crystal KDP (Potassium dihydrogen phosphate) with $\chi^{(2)}$ optical nonlinearity.

$$\omega_p = \omega_1 + \omega_2 = \omega_i + \omega_s \qquad i = idler$$
$$k_p = k_1 + k_2 = k_i + k_s \qquad s = signal$$

Phases of the corresponding wavefunctions match

$$|\Psi\rangle = \int_0^{\omega_p} \phi(\omega_1, \omega_0 - \omega_1) \ |\omega_1\rangle|\omega_0 - \omega_1\rangle \, d\omega_1$$

2 photons leave the crystal with the same polarization but orthogonal to polarization of the pump beam.

Type II:
The two emitted photon pairs have orthogonal polarizations. This allows polarizations to be used for entanglement achieving hyper-entanglement.

$$|\Psi\rangle = \frac{1}{2}\int_0^{\omega_p} \phi(\omega, \omega_0 - \omega) \ |\omega\rangle|\omega_0 - \omega\rangle \cdot \left[(|k_1\rangle|k_1'\rangle + e^{i\phi}|k_2\rangle|k_2'\rangle)(|e\rangle|0\rangle + |0\rangle e\rangle)\right] d\omega$$

where
$e = extraordinary\ axes\ of\ nonlinear\ crystal$
$o = ordinary\ axes\ of\ nonlinear\ crystal$

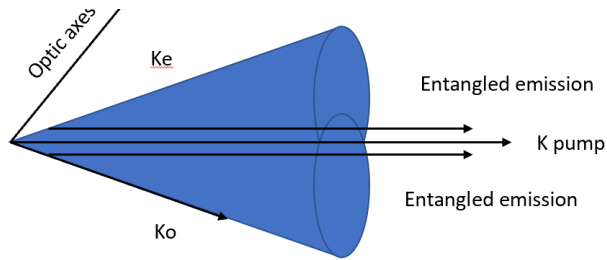

Figure 4. Focusing Optics for Muxing 4 OAM Modes

High-Intensity type II phase-matched SPDC 2-photon entanglement with no extra beam splitters.

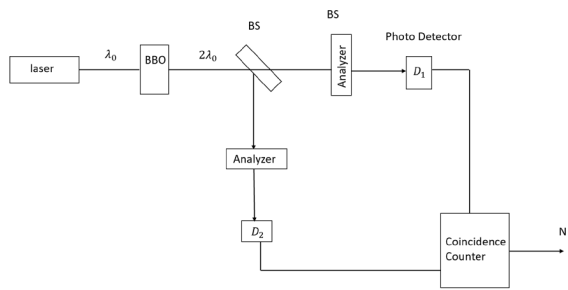

Figure 5. Focusing Optics for Muxing 4 OAM Modes

One can generate all four Bell states
$|\Psi^\pm\rangle = \frac{1}{2}[(|HV\rangle \pm |VH\rangle)], |\Phi^\pm\rangle = \frac{1}{2}[(|HH\rangle \pm |VV\rangle)]$

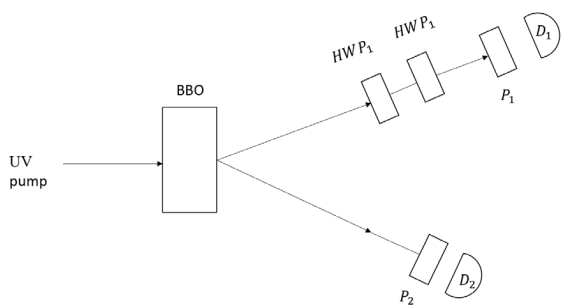

Figure 6. Focusing Optics for Muxing 4 OAM Modes

New Two-Crystal ultra-bright 2-photon entanglement

Line polarized 45° pump

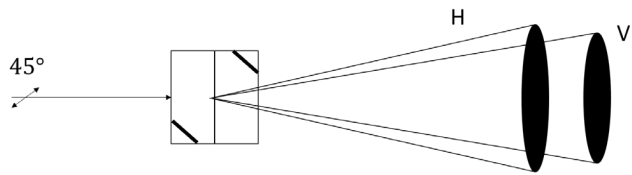

Figure 7. Focusing Optics for Muxing 4 OAM Modes

2-photons have $|\Psi\rangle = \frac{1}{\sqrt{2}}(|H\rangle|H\rangle + e^{i\phi}|V\rangle|V\rangle)$

where $\phi$ = adjusted by crystal tilt

One can measure Polarization Mode Dispersion (PMD) which is the difference in propagation rate between two polarizations in a birefringent medium. We can also generate entangled short and long distance

$|\Psi\rangle = \frac{1}{\sqrt{2}}(|S\rangle_p|S\rangle_p + e^{i\phi}|L\rangle_p|L\rangle_p)$    $p = pump$
2-photon state pairs

$|\Psi\rangle = a|HH\rangle + b|HV\rangle + c|VH\rangle + d|VV\rangle$

Entanglement happens because of indistinguishability of HH and VV pairs prior to polarization measurements

$$|\Psi\rangle = \frac{1}{\sqrt{2}}(|HH\rangle + e^{i\phi}|V\rangle|V\rangle)$$

We get this if we cannot tell from which crystal a certain photon pair emerge.

The compensation crystal can be tilted so that we can get $\phi = 0$ (relative) Producing

$$|\Psi\rangle = \frac{1}{\sqrt{2}}(|HH\rangle + |V\rangle|V\rangle)$$

In a two-photon system that are Gaussian input signals

$$|\Psi_{in}\rangle = (\alpha|H\rangle|V\rangle + \beta e^{i\phi}|V\rangle|H\rangle) \otimes |0\rangle|0\rangle$$

$$\alpha^2 + \beta^2 = 1$$

H : Horizontal polarization
V : Vertical polarization

The photons can be transferred to a LG mode by SLM based on the polarization of photons.

$|\Phi\rangle_{12345} = \frac{1}{\sqrt{2}}[|H_1\rangle|H_2\rangle|H_3\rangle|H_4\rangle|H_5\rangle + |V_1\rangle|V_2\rangle|V_3\rangle|V_4\rangle|V_5\rangle]$    $\_\rangle + e^{i\phi}\beta|l_-\rangle|l_+\rangle$

If we put a polarizer after this stage, we get

$$|\Psi_{out}\rangle_p = |D\rangle|D\rangle \otimes [\alpha|l_+\rangle|l_-\rangle + e^{i\phi}\beta|l_-\rangle|l_+\rangle]$$

Multi-photon Entanglement

3-particle GHZ-Entanglement
Where GHZ: Greenberger-Horne-Zeilinger

3-GHZ-states are generalized Bell-states

$$|\Phi^\pm\rangle = \frac{1}{2}[|H\rangle|H\rangle|H\rangle \pm |V\rangle|V\rangle|V\rangle]$$
$$|\Psi_1^\pm\rangle = \frac{1}{2}[|V\rangle|H\rangle|H\rangle \pm |H\rangle|V\rangle|V\rangle]$$
$$|\Psi_2^\pm\rangle = \frac{1}{2}[|H\rangle|V\rangle|H\rangle \pm |V\rangle|H\rangle|V\rangle]$$
$$|\Psi_3^\pm\rangle = \frac{1}{2}[|H\rangle|H\rangle|V\rangle \pm |V\rangle|V\rangle|H\rangle]$$

$$|\Psi^\pm\rangle_{23} = \frac{1}{\sqrt{2}}[|H\rangle|V\rangle \pm |V\rangle|H\rangle]$$
$$|\Phi^\pm\rangle_{23} = \frac{1}{\sqrt{2}}[|H\rangle|H\rangle \pm |V\rangle|V\rangle]$$

Photon 2 and 3 are measured, projecting their state onto one of the Bell states

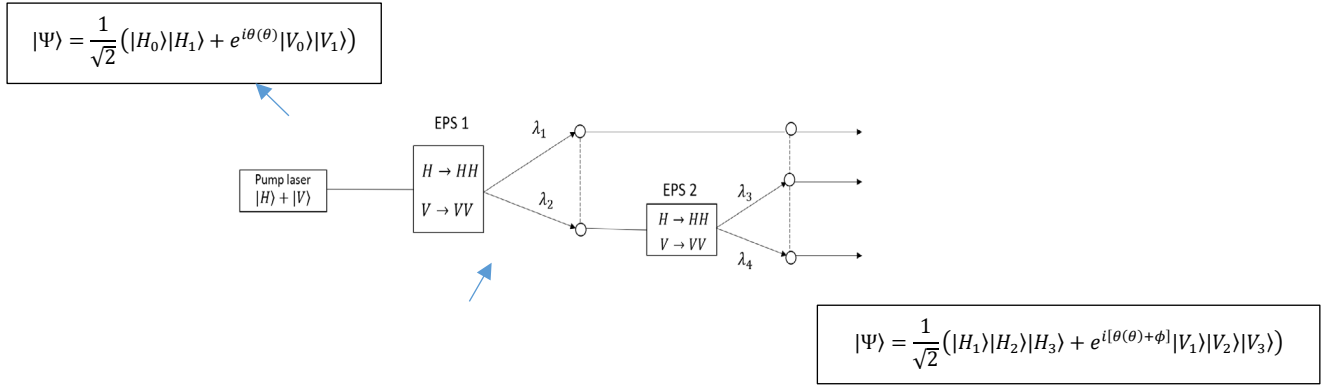

Figure 8. Focusing Optics for Muxing 4 OAM Modes

$\sigma_x, \sigma_y = -i|H\rangle\langle V| - i|V\rangle\langle H|$
$\sigma_z = |H\rangle\langle H| - |V\rangle\langle V|$
EPS: entangled photon source

Five-photon Entanglement

Here we can use two-entangled photon pairs to generate a four-photon entangled sate, which is then combined with a single photon state to achieve 5-photon entanglement.

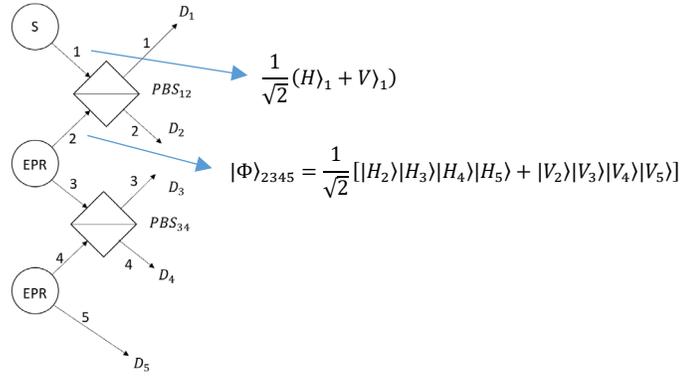

Figure 9. Focusing Optics for Muxing 4 OAM Modes

Entangled states are in Bell-states are:

$$|\Phi^+\rangle_{23} = \frac{1}{\sqrt{2}}[|H\rangle_2|H\rangle_3 + |V\rangle_2|V\rangle_3]$$
$$|\Phi^-\rangle_{23} = \frac{1}{\sqrt{2}}[|H\rangle_2|H\rangle_3 - |V\rangle_2|V\rangle_3]$$
$$|\Psi^+\rangle_{23} = \frac{1}{\sqrt{2}}[|H\rangle_2|V\rangle_3 + |V\rangle_2|H\rangle_3]$$
$$|\Psi^-\rangle_{23} = \frac{1}{\sqrt{2}}[|H\rangle_2|V\rangle_3 - |V\rangle_2|H\rangle_3]$$

Similarly for $|\Phi^+\rangle_{45}, |\Phi^-\rangle_{45}, |\Psi^+\rangle_{45}, |\Psi^-\rangle_{45}$ with index 45

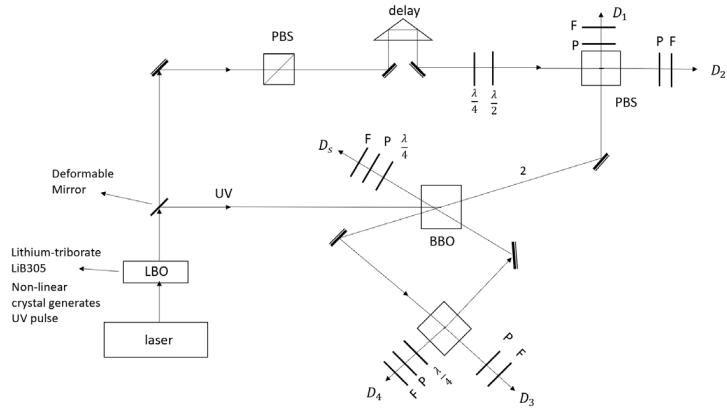

Figure 10. Focusing Optics for Muxing 4 OAM Modes

Six-photon GHZ-state

$$|G_6\rangle = \frac{1}{\sqrt{2}}[|H_1\rangle|H_2\rangle|H_3\rangle|H_4\rangle|H_5\rangle|H_6\rangle + |V_1\rangle|V_2\rangle|V_3\rangle|V_4\rangle|V_5\rangle|V_6\rangle]$$

Applying Hadamard to photon 4

$\mathbb{H}|H_4\rangle = |+_4\rangle$
$\mathbb{H}|V_4\rangle = |-_4\rangle$

Four-photon GHZ state (combine 2 and 3 photons)

$$\frac{1}{\sqrt{2}}[|H_1\rangle|H_2\rangle|H_3\rangle|+_4\rangle + |V_1\rangle|V_2\rangle|V_3\rangle|-_4\rangle]$$

Where

$$|+\rangle = \frac{1}{\sqrt{2}}[|H\rangle + |V\rangle]$$
$$|-\rangle = \frac{1}{\sqrt{2}}[|H\rangle - |V\rangle]$$

Combine photon 4 and 5, we have six-photon cluster state

$$|C_6\rangle = \frac{1}{2}[|H_1\rangle|H_2\rangle|H_3\rangle|H_4\rangle|H_5\rangle|H_6\rangle + |H_1\rangle|H_2\rangle|H_3\rangle|V_4\rangle|V_5\rangle|V_6\rangle + |V_1\rangle|V_2\rangle|V_3\rangle|H_4\rangle|H_5\rangle|H_6\rangle - |V_1\rangle|V_2\rangle|V_3\rangle|V_4\rangle|V_5\rangle|V_6\rangle]$$

Six-photon GHZ

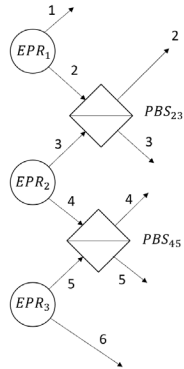

Figure 11.  Focusing Optics for Muxing 4 OAM Modes

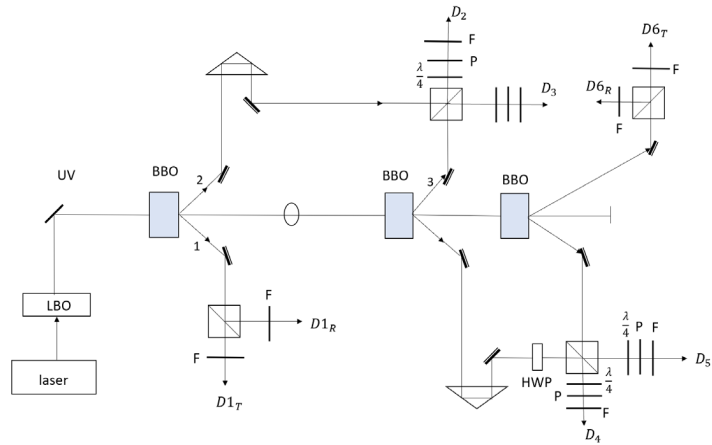

Figure 12.  Focusing Optics for Muxing 4 OAM Modes

Ten-photons

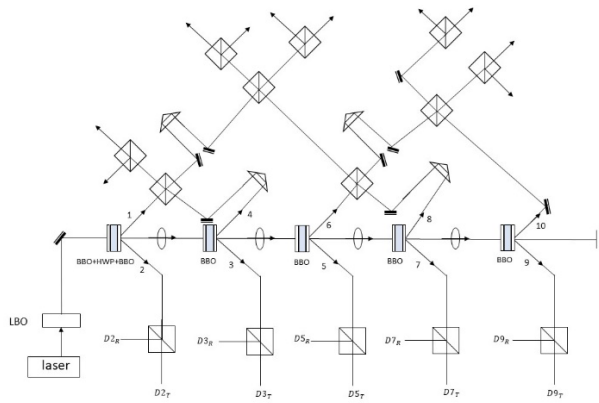

Figure 13.  Focusing Optics for Muxing 4 OAM Modes

Ten Photons with ten helicities

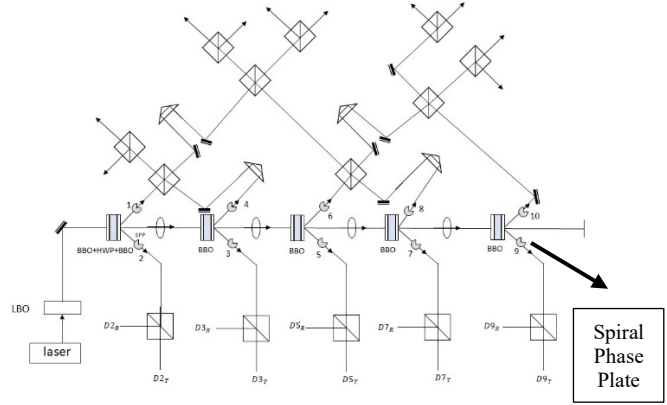

Figure 14. Focusing Optics for Muxing 4 OAM Modes

C-NOT operation with multi-photon hyper-entangled qudits with polarization as a control bit

$$(\alpha|H\rangle + \beta|V\rangle)|l_+\rangle \rightarrow (\alpha|H\rangle|l_+\rangle + \beta|V\rangle|l_-\rangle)$$

Steps

$$(\alpha|H\rangle + \beta|V\rangle)\left(\frac{\Pi|l_+\rangle + \Pi|l_-\rangle}{\sqrt{2}}\right)$$

Where

$$\Pi|l_+\rangle = |l_{+1}\rangle|l_{+2}\rangle|l_{+3}\rangle|l_{+4}\rangle \ldots$$
$$\Pi|l_-\rangle = |l_{-1}\rangle|l_{-2}\rangle|l_{-3}\rangle|l_{-4}\rangle \ldots$$

Then $(\alpha|H\rangle + \beta|V\rangle)\left(\frac{\Pi|l_+\rangle}{\sqrt{2}}\right)$ gives

$$\left(\alpha|H\rangle e^{-i\frac{\pi}{4}} + \beta|V\rangle e^{i\frac{\pi}{4}}\right)\left(\frac{\Pi|l_+\rangle}{\sqrt{2}}\right)$$

and $(\alpha|H\rangle + \beta|V\rangle)\left(\frac{\Pi|l_-\rangle}{\sqrt{2}}\right)$ gives

$$\left(\alpha|H\rangle e^{i\frac{\pi}{4}} + \beta|V\rangle e^{-i\frac{\pi}{4}}\right)\left(\frac{\Pi|l_-\rangle}{\sqrt{2}}\right)$$

$$e^{i\frac{\pi}{4}}\alpha|H\rangle\left[\frac{\Pi|l_+\rangle - i\Pi|l_-\rangle}{\sqrt{2}}\right] + e^{-i\frac{\pi}{4}}\beta|V\rangle\left[\frac{\Pi|l_+\rangle + i\Pi|l_-\rangle}{\sqrt{2}}\right]$$

then

$$\alpha|H\rangle\Pi|l_+\rangle + \beta|V\rangle\Pi|l_-\rangle$$

$$\alpha|H\rangle|U\rangle\Pi|l_+\rangle + \beta|V\rangle|D\rangle\Pi|l_-\rangle$$
$$\alpha|H\rangle|U\rangle|S\rangle\Pi|\lambda_i\rangle\Pi|l_+\rangle + \beta|V\rangle|D\rangle|L\rangle\Pi|\lambda_j\rangle\Pi|l_-\rangle$$

Where degrees of freedom are
U: up beam path

D: down beam path  
H: horizonal polarization  
V: vertical polarization  
$l_-$ , $l_+$ = negative and positive helicities  
S: short or early  
L: long or late  
$\lambda_i, \lambda_j$: resolvable wavelengths

## VI. DEGREES OF FREEDOM

Any two-level quantum system can represent a qubit. Thus, a typical qubit could be a spin-1/2. But experiments with spin-1/2 systems are difficult. Fortunately, photons possess a ready and easily controllable qubit degree of freedom. Spin in quantum electrodynamics has an analog in classical electrodynamics called polarization. However, there are also other methods to make photons carry qubits such as OAM states of photons. The following is a list of few degrees of freedom that can be used for quantum computing:

<u>Polarization qubits:</u> The most commonly used photonic qubits are realized using polarization. In this case arbitrary qubit states can be

For two-photon polarization

HV Linear $[|H\rangle, |V\rangle]$ 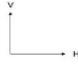

$D\overline{D}$ Diagonal $\left[\frac{1}{\sqrt{2}}(|H\rangle + |V\rangle), \frac{1}{2}(-|H\rangle + |V\rangle)\right]$ 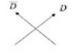

$RL$ Circular $\left[\frac{1}{\sqrt{2}}(|H\rangle - i|V\rangle), \frac{1}{2}(|H\rangle - i|V\rangle)\right]$ 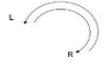

$$|\Psi\rangle_2 = \alpha|H\rangle + \beta|V\rangle$$

where H and V stand for horizontal and vertical polarizations respectively. The advantage of using polarization qubits comes from the fact that they can easily be created and manipulated with high precision (at 99% level) by simple linear-optical elements such as polarizing beam splitters (PBS), polarizers and waveplates.

<u>Spatial qubits:</u> A single photon can also appear in two different spatial modes or paths (say up and down paths). The general state reads

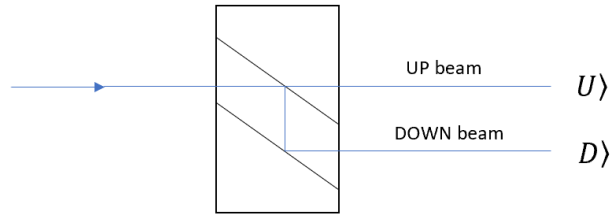

Figure 15. Focusing Optics for Muxing 4 OAM Modes

$$|\Psi\rangle_2 = \alpha|D\rangle + \beta|U\rangle$$

Where *U*: up beam path and *D*: down beam path

This can be achieved if a single photon exits a beam splitter (BS), with two output modes D and U. Any state of spatial qubits can be prepared by using suitable phase shifters and beam splitters. A disadvantage of using spatial qubits is that the coherence between two output modes D and U is sensitive to the relative phase for the two paths and this is difficult to control in long-distance cases.

Time-bin qubits: For a more robust long-distance transmission of quantum information, one can use time-bin qubits. The computational basis consists of two states which are of the same spectral shape, but time shifted by much more than the coherence time: early and late. Any state of time-bin qubits can be realized with a single-photon pulse sent through an unbalanced Mach-Zehnder interferometer.

$$|\Psi\rangle = \frac{1}{\sqrt{2}}\left(|S\rangle_p|S\rangle_p + e^{i\phi}|L\rangle_p|L\rangle_p\right) \qquad p = pump$$

Frequency or Wavelength-bin qudits: This is a degree of freedom that uses resolved frequencies or wavelength of photonic beam as orthogonal basis.

$$|\Psi\rangle_n = \alpha|\lambda_i\rangle + \beta|\lambda_j\rangle$$

OAM qudits: This is a degree of freedom that uses resolved OAM modes of photonic beam. However, unlike the two-state polarization, OAM can have many orthogonal resolvable modes and they can lend themselves to creating qudits rather than qubits.

$$|\Psi\rangle_n = \alpha|l_+\rangle + \beta|l_-\rangle$$

## VII. OAM BEAMS MULTIPLEXING AND DEMULTIPLEXING

Many approaches for creating OAM beams have been proposed and demonstrated. One can obtain a single or multiple OAM beams directly from the output of a laser cavity or can convert a fundamental Gaussian beam into an OAM beam outside a cavity. The converter could be a spiral phase plate, diffractive phase holograms, metamaterials, cylindrical lens pairs, q-plates, fiber gratings, or couplers. There are also different ways to detect an OAM beam, such as using a converter that creates a conjugate helical phase or by using a plasmonic detector.

One of the benefits of OAM is that multiple coaxially propagating OAM beams with different l states provide additional data carriers, as they can be separated based only on the twisting wave front. Hence, one of the critical techniques is the efficient multiplexing/demultiplexing of OAM beams of different l states, such that each carries an independent data channel and all beams can be transmitted and received using a single aperture pair. Several multiplexing and demultiplexing techniques have been demonstrated, including the use of an inverse helical phase hologram to convert the OAM into a Gaussian-like beam, a mode sorter [10], free-space interferometers, a photonic integrated circuit, and q-plates. Some of these techniques are briefly described below.

- Beam Splitter and Inverse Phase Hologram

A straightforward way of multiplexing different beams is to use cascaded beam splitters. Each beam splitter can coaxially multiplex two beams that are properly aligned, and cascaded N beam splitters can multiplex N + 1 independent OAM beams at most. Similarly, at the receiver end, the multiplexed beams are divided into four copies. To demultiplex the data channel on one of the beams, a phase hologram with a spiral charge is applied to all the multiplexed beams. As a result, the helical phase on the target beam is removed, and this beam will eventually evolve into a Gaussian-like beam. The Gaussian-like beam can be isolated from the other OAM beams, which still have helical phase fronts, by using a spatial mode filter, e.g., a single-mode fiber (SMF) at the focal point of a lens will couple the power only of the Gaussian mode due to the mode-matching constraints. Accordingly, each of the multiplexed beams can be demultiplexed by changing the spiral phase hologram. Although the power loss incurred by the beam splitters and the spatial mode filter makes this method quite power inefficient, it was frequently used in the initial laboratory demonstrations of OAM multiplexing/demultiplexing due to its simplicity and reconfigurability provided by the programmable SLMs.

-Optical Geometrical Transformation-Based Method

More power-efficient multiplexing and demultiplexing of OAM beams are achieved by using an OAM mode sorter [10-13]. This mode sorter usually comprises three optical elements, namely, a transformer, a corrector, and a lens. The transformer performs a geometrical transformation of the input beam from log-polar coordinates to Cartesian coordinates, such that the position (x; y) in the input plane is mapped to a new position (u; v) in the output plane. The corrector compensates for phase errors and ensures that the transformed beam is collimated. Considering an input OAM beam with ring-shaped beam profile, it can be unfolded and mapped into a rectangular-shaped plane wave with a tilted phase front. Similarly, multiple OAM beams having different l states will be transformed into a series of plane waves, each with a different phase tilt. A lens focuses these tilted plane waves into spatially separated spots in the focal plane such that all OAM beams are simultaneously demultiplexed. Since the transformation is reciprocal, the mode sorter can also be used in reverse to function as an OAM generator and multiplexer. A Gaussian beam array placed in the focal plane of the lens is converted into superimposed plane waves with different tilts. These beams then pass through the corrector and the transformer to produce properly multiplexed OAM beams [13].

*Proof-of-Concept Link demonstration*

Initial demonstrations of using OAM multiplexing for optical communications include free-space links using a Gaussian beam and an OAM beam encoded with on–off keying data. We have demonstrated, the mmultiplexing /demultiplexing of four different OAM beams in a free-space data link. Four monochromatic Gaussian beams each carrying an independent 50.8 Gbps (4 × 12.7 Gbps) 16-QAM signal were prepared from an IQ modulator and free-space collimators.

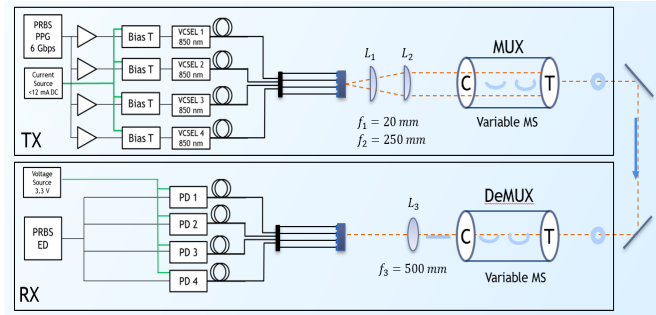

Figure 16. Muxing and de-Muxing of 4 OAM Modes

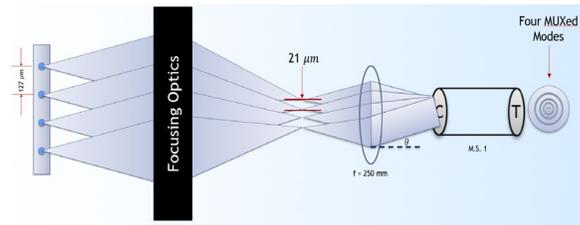

Figure 17. Focusing Optics for Muxing 4 OAM Modes

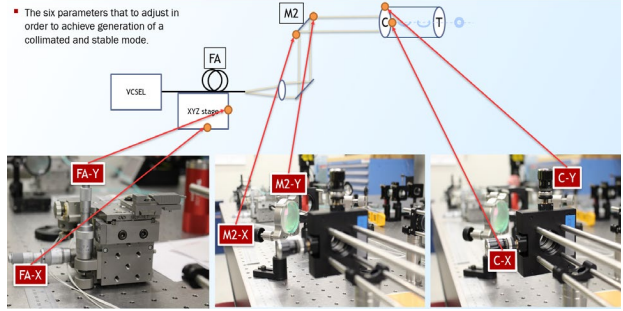

Figure 18. Optical Components for Muxing 4 OAM Modes

Combination qudits (not qubits): It is indeed possible to combine all these degrees of freedom to achieve a higher qudit of higher order radix as follows:

$$\begin{pmatrix}H\\V\end{pmatrix} \otimes \begin{pmatrix}U\\D\end{pmatrix} \otimes \begin{pmatrix}l_{+1}\\l_{+2}\\\vdots\\l_{+5}\\l_{-1}\\l_{-2}\\\vdots\\l_{-5}\end{pmatrix} = \begin{pmatrix}HU\\HD\\VU\\VD\end{pmatrix} \otimes \begin{pmatrix}l_{+1}\\l_{+2}\\\vdots\\l_{+5}\\l_{-1}\\l_{-2}\\\vdots\\l_{-5}\end{pmatrix} = \begin{pmatrix}HUl_{+1}\\HUl_{+2}\\\vdots\\\vdots\\VDl_{-5}\end{pmatrix}$$

We can even combine wavelength

$$\alpha |H\rangle|U\rangle \prod |\lambda_i\rangle \prod |l_+\rangle + \beta |V\rangle|D\rangle \prod |\lambda_j\rangle \prod |l_-\rangle$$

and time-bins with early and late or short and long

$$\alpha |H\rangle|U\rangle|S\rangle \prod |\lambda_i\rangle \prod |l_+\rangle + \beta |V\rangle|D\rangle|L\rangle \prod |\lambda_j\rangle \prod |l_-\rangle$$

If we compare this to Google's quantum supremacy that was achieved with 53 qubits, we can see how many qudits can be used to cover the same computational state-space dimension using only 10 photons. It is possible to achieve quantum supremacy by using 10 photons with radix-40 which is possible to achieve based on the prescription above because a ten-photon entanglement has already been achieved. The calculation is as follows:

$$2^{53} = x^{10}$$

$$53 \log_{10} 2 = 10 \log_{10} x$$

$$\log_{10} x = 1.595$$

$$x = 39.36 \qquad x \approx 40$$

## VIII. A New Performance Measure for Quantum Computing

Recent quantum computing has been entirely about how many qubits one can achieve. If one achieves 53 qubits, then you can claim quantum supremacy. However, efforts now focused on managing systems with several tens of qubits [54-56]. In such noisy environment of intermediate-scale quantum (NISQ) systems [57], performance of gates could not predict the behavior of the system. Methods such as randomized benchmarking [58], state and process tomography [59], and gateset tomography [60] can be used for measuring the performance of operations on a few qubits, yet they neglect to account for errors from interactions with qubits [61,63]. Given a system like this, where individual gate operations have been independently adjusted and verified, it is not easy to measure the degree to which the system performs as a general purpose universal quantum computer. Here we introduce a new performance measure which is different than IBM's quantum volume. This new measure is also strongly linked to gate error rates and is affected by underlying qubit connectivity and gate parallelism. It can also be improved by achieving the limit of large numbers of controlled, highly coherent, connected, and generically programmable qubits are manipulated within a circuit. This new measure has to do with computational state-space of dimension, high-fidelity operations, high connectivity, large calibrated gate sets, and circuit rewriting toolchains. This new measure which we call quantum capacity is a practical way to measure and compare progress toward improved system structure of a universal quantum computer. Note that it is possible to achieve the same computational state-space using higher number of qubits or lower number of qubits with higher radix. For example, it is possible to achieve the same 53 qubit computational state-space dimension by using only 10 photons of radix-40 and 10-photon entanglement has already been achieved.

## Conclusion

The approach presented in this paper can achieve a convergence of an advanced FSO system for communications, QKD, quantum computing and quantum teleportation with qudits that can perform beyond what was achieved with the Chinese Micius quantum satellite. Based on what Raytheon has already achieved, its novel system can initially deliver a low cost 10 Gbps FSO link scalable to 100 Gbps without using OAM methods. Since Raytheon's system uses Gaussian beams to deliver such throughputs, it is possible to increase the throughput using NxGen's system leveraging OAM muxing and demuxing as a new dimension to increase link throughput. The collaboration between Raytheon and NxGen can potentially increase Raytheon's system throughput from an initial 4× with 4 OAM modes, 8× with 4 OAM modes and 2 polarizations states, to 24× with 12 OAM modes and 2 polarization states reaching an output of 2.4 Tbps using a single optical frequency.